\documentclass{emulateapj}

\usepackage{apjfonts}
\usepackage{natbib}

\shorttitle{MASSIVE GLOBULAR CLUSTERS IN NGC~5128} 
\shortauthors{MARTINI \& HO} 

\slugcomment{ApJ accepted [30 March 2004]}

\newcommand{\cena}{NGC~5128} 

\newcommand{\msun}{M$_{\odot}$}
\newcommand{\hst}{{\it HST}}
\newcommand{\eg}{{\rm e.g.}}

\newcommand{\kms}{km s$^{-1}$}

\begin{document}

\title{A Population of Massive Globular Clusters in NGC~5128} 

\author{Paul Martini} 

\affil{Harvard-Smithsonian Center for Astrophysics; 
60 Garden Street, MS20; Cambridge, MA 02138; pmartini@cfa.harvard.edu}

\and

\author{Luis C. Ho} 

\affil{Observatories of the Carnegie Institution of Washington; 
813 Santa Barbara Street; Pasadena, CA 91101; lho@ociw.edu}

\begin{abstract}

We present velocity dispersion measurements of 14 globular clusters in 
NGC~5128 (Centarus~A) obtained with the MIKE echelle spectrograph 
on the 6.5m Magellan Clay telescope. These clusters are among the 
most luminous globular clusters in \cena\ and have velocity dispersions 
comparable to the most massive clusters known in the Local Group, ranging 
from 10 -- 30 \kms. 
We describe in detail our cross-correlation measurements, as well as 
simulations to quantify the uncertainties. 
These 14 globular clusters are the brightest \cena\ globular clusters with 
surface photometry and structural parameters measured from the {\it Hubble 
Space Telescope}. 
We have used these measurements to derive masses and mass-to-light ratios for 
all of these clusters and establish that the fundamental plane relations for 
globular clusters extend to an order of magnitude higher mass than in the 
Local Group. 
The mean mass-to-light ratio for the \cena\ clusters is $\sim 3 \pm 1$, higher 
than measurements for all but the most massive Local Group clusters. 
These massive clusters begin to bridge the mass gap between the most massive 
star clusters and the lowest-mass galaxies. 
We find that the properties of \cena\ globular clusters overlap quite well 
with the central properties of nucleated dwarf galaxies and 
ultracompact dwarf galaxies. 
As six of these clusters also show evidence for extratidal light, we 
hypothesize that at least some of these massive clusters are the nuclei of 
tidally stripped dwarfs. 

\end{abstract}


\section{Introduction} \label{sec:intro}

Globular clusters provide valuable snapshots of the formation history of 
galaxies and their large sizes and luminosities make them the most readily 
observable sub-galactic constituents. 
In addition, globular clusters exhibit surprisingly uniform properties 
that suggests a common formation mechanism. 
They are well-fit by single-mass, isotropic King models \citep{king66}, 
which describe clusters in terms of scale radii, central surface 
brightness, and core velocity dispersion. 
Detailed studies of globular clusters in our Galaxy have shown that in 
fact they only inhabit a narrow range of the parameter space available 
to King models \citep{djorgovski95,mclaughlin00} and other globular 
cluster systems in the Local Group also appear to approximately follow the 
same relations \citep{djorgovski97,dubath97a,dubath97b,barmby02,larsen02}. 
These parameter correlations trace a globular cluster fundamental plane 
that is analogous to, but distinct from, the fundamental plane for 
elliptical galaxies
\citep{dressler87,djorgovski87,bender92,burstein97}. 

Globular cluster studies that include internal kinematics have been 
confined to the Local Group due to the faint apparent magnitudes of 
more distant extragalactic globular clusters. These studies have therefore 
only included the globular cluster systems of spiral and dwarf galaxies 
and not the globular cluster systems of large ellipticals. 
Yet the globular cluster systems of ellipticals are a particularly 
interesting regime as they probe both a new morphological type and one likely 
to have exhibited a different and more complex formation history. 
The globular cluster systems of elliptical galaxies, as well as many spirals 
such as our own, have bimodal color distributions suggestive of multiple 
episodes of formation \citep[\eg][]{kundu01,larsen01a}. 
Models for the formation of these globular cluster systems posit that one of 
these populations may be the intrinsic population of the galaxy and 
subsequent mergers resulted in the second population 
either as the result of a new episode of globular 
cluster formation \citep*{schweizer87,ashman92,forbes97} or accretion of 
globular cluster systems from other galaxies, including accretion of 
globular clusters by tidal stripping from other members of a cluster of 
galaxies \citep*{cote98}.  

NGC~5128 (Centarus~A), as the nearest large elliptical 
galaxy, is arguably the best source for extending detailed globular cluster 
studies outside of the Local Group. 
While \cena\ is the central galaxy of a large group, rather 
than a giant elliptical in a cluster, it likely had a similar formation 
history to its larger cousins. The most relevant similarity is the strong 
evidence for a recent, gas-rich merger 
\citep[for a recent review of \cena\ see][]{israel98}. 
Estimates of the size of the \cena\ globular cluster population suggest 
that it has a total of $\sim 2000$ clusters, approximately a factor of 
3 more than the entire Local Group \citep{harris84}. 
Simple scaling arguments suggest that \cena\ should possess a number of 
extremely massive globular clusters 
and therefore is not only a good target for study of the globular cluster 
system of an elliptical galaxy, but also for study of how well the 
fundamental plane relations established locally apply to more massive 
globular clusters. 
A recent photometric and spectroscopic study of \cena\ by 
\citet*{peng04a,peng04b} concluded that the metal-rich globular clusters
may have a mean age of $5^{+3}_{-2}$ Gyr, while an analysis of their 
photometric data yields a metallicity range of 
$-2.0$ through $+0.3$  \citep{yi04} 

The most massive globular clusters can also be used to explore connections  
between the formation processes for star clusters and galaxies. 
While fundamental plane studies \citep[\eg][]{burstein97} clearly illustrate 
a significant mass gap between the most massive Galactic globular clusters and 
the least massive dwarf galaxies, there have been encroachments into this gap 
from both sides. For many years, studies have speculated that at least some 
globular clusters may be the remains of tidally-stripped dwarf galaxy nuclei 
\citep{zinnecker88,freeman93,bassino94}. 
Two of the most massive globular clusters in the Local Group, $\omega$Cen in 
our Galaxy and G1 in M31, have been interpreted as the nuclei of tidally 
stripped dwarfs \citep{meylan01,gnedin02,bekki03b}. 
From the galaxy side, recent studies of nucleated dwarf galaxies in the Virgo 
cluster \citep*{geha02} and ultracompact dwarf galaxies in the Fornax cluster 
\citep{drinkwater03} show some similarities between these least-massive 
galaxies and the most massive globular clusters. This mass gap may thus 
reflect the scarcity of the most massive globular clusters and the difficulty 
of kinematic measurements for the least massive, lowest surface-brightness 
dwarf galaxies, rather than a physical separation. 

In this paper we present velocity dispersion measurements for 14
globular clusters in \cena. These globular clusters were selected from 
the {\it Hubble Space Telescope} (\hst) study of \citet{harris02} and 
therefore have well-measured 
structural parameters. These data are combined to estimate masses for 
these clusters, masses that are among the largest known for any star clusters 
and comparable to the nuclei of the lowest-mass galaxies. In the 
next sections we describe the observations, data processing, and velocity 
dispersion measurements. Analysis of these measurements is described 
in \S5 and the potential link between star clusters and galaxies is 
explored in \S6. Our results are summarized in the final section. 
Throughout this paper we adopt the distance of $3.84 \pm 0.35$ Mpc for 
\cena\ determined by \citet{rejkuba04} from the brightness of the tip of the 
red giant branch and the Mira period-luminosity relation. 

\section{Observations}  \label{sec:obs}

Spectra of 14 globular clusters in \cena\ were obtained in the 
course of seven nights in March 2003 with the MIKE echelle spectrograph 
\citep{bernstein03} and the 6.5m Magellan Clay telescope at Las 
Campanas Observatory (see Table~\ref{tbl:obs}). 
MIKE is a double echelle spectrograph capable of resolution $R = 19,000$ on 
the red side when used with the $1''$ slit selected for these observations. 
This corresponds to a velocity resolution FWHM of 15.8 \kms\ or 
$\sigma = 6.2$ \kms. 
MIKE was located at the East Nasmyth port but not directly attached to the 
instrument rotator. The advantage of this configuration is that the 
instrument is gravity invariant and therefore potential calibration 
difficulties due to instrument flexure are completely avoided. The 
disadvantage of this configuration is that targets could not be observed at 
the parallactic angle and that they rotate with respect to the spectrograph 
slit. In any event, this configuration was the only one available at the time. 
To minimize lost light due to atmospheric dispersion, the position angle of 
the spectrograph was set to correspond to the parallactic angle for an object 
at airmass 1.3; below airmass $\sim 1.4$ the atmospheric dispersion correction 
is effectively negligible and most of our observations took place in this 
airmass range. Any lost light due to atmospheric dispersion is unlikely to 
significantly impact our velocity dispersion measurements as the velocity 
dispersion gradient of the clusters is not large. The 
rotation of the slit also has a negligible impact as all of these globular 
clusters are fairly round \citep{harris02}. 

\begin{center}
\begin{deluxetable}{lcccc}
\tablecolumns{5}
\tablecaption{Globular Cluster Observations and Measurements\label{tbl:obs}}
\tablehead{
\colhead{ID} &
\colhead{$V$ [mag]} &
\colhead{Exptime [s]} &
\colhead{SNR} & 
\colhead{$\sigma$ [\kms]} \\
}
\startdata
C2  . . . . . . . . . . & 18.33 & 10800 & 11 & $ 14.2 \pm 2.0 $  \\
C7  . . . . . . . . . . & 17.10 & 5400  & 16 & $ 22.4 \pm 2.1 $  \\
C11 . . . . . . . . . . & 17.70 & 9000  & 18 & $ 17.7 \pm 1.9 $  \\
C17 . . . . . . . . . . & 17.61 & 7200  & 19 & $ 18.9 \pm 2.0 $  \\
C21 . . . . . . . . . . & 17.77 & 10800 & 12 & $ 20.8 \pm 1.9 $  \\
C22 . . . . . . . . . . & 18.14 & 10800 & 15 & $ 19.1 \pm 2.0 $  \\
C23 . . . . . . . . . . & 17.19 & 7200  & 25 & $ 31.4 \pm 2.6 $  \\
C25 . . . . . . . . . . & 18.33 & 12600 &  5 & $ 12.2 \pm 2.0 $  \\
C29 . . . . . . . . . . & 17.94 & 14400 & 12 & $ 16.1 \pm 2.1 $  \\
C31 . . . . . . . . . . & 18.37 & 12600 &  8 & $ 15.0 \pm 2.1 $  \\
C32 . . . . . . . . . . & 18.31 & 10800 &  7 & $ 15.8 \pm 1.4 $  \\
C37 . . . . . . . . . . & 18.34 & 10800 & 10 & $ 13.5 \pm 1.6 $  \\
C41 . . . . . . . . . . & 18.56 & 19800 &  8 & $  9.6 \pm 2.0 $  \\
C44 . . . . . . . . . . & 18.61 & 18000 &  8 & $  9.1 \pm 2.0 $  \\
\enddata
\tablecomments{
Present sample of \cena\ globular clusters. Cols.~1 and 2 list the cluster 
ID and apparent $V$ magnitude from \citet{peng04a} and col.~3 our exposure 
times. The mean signal-to-noise ratio over the range 5000 -- 6800 \AA\ 
per $0.2$ \AA\ is listed in col.~4. The measured velocity 
dispersions and uncertainties, described in \S\ref{sec:losvd}, are listed in 
column 5. 
} 
\end{deluxetable}
\end{center}

While MIKE was used to obtain both red and blue data, sufficient 
signal-to-noise ratio (SNR) observations of these (intrinsically red) clusters 
were only obtained on the red side. 
To enhance the SNR, the observations were binned $2 \times 2$, leading to 
$0.286''$ pixel$^{-1}$ and a velocity scale of 4.2 \kms pixel$^{-1}$. 
Table~\ref{tbl:obs} lists all 14 clusters according to their 
identification in \citet{harris02}, along with their apparent $V$ magnitude 
from \citet{peng04a}, the total integration time, and the average 
SNR per pixel over the wavelength range 5000 -- 6800 \AA\ when binned to 
0.2 \AA\ pixel$^{-1}$. The total integration time 
listed is the sum of an integer number of 1800 s exposures. This exposure time 
was chosen to facilitate cosmic ray removal, yet also avoid a significant 
contribution from detector noise. 
Spectra of 16 velocity dispersion templates were also obtained, ranging 
in spectral type from G through M stars and luminosity classes from IV 
through I.
These templates were observed in a 
similar manner to the globular clusters. Typically three integrations of a 
few seconds each were obtained for each template. Arc lamps for wavelength 
calibration were obtained before or after the observations of each target. 
Several sets of flatfield exposures using a quartz lamp were also obtained 
over the course of the seven-night observing run.

\section{Data Processing}  \label{sec:data}

The data were processed with a combination of tasks from the 
IRAF\footnote{IRAF is distributed by the National Optical Astronomy
Observatories, which are operated by the Association of Universities for 
Research in Astronomy, Inc., under cooperative agreement with the National 
Science Foundation.} echelle package and the sky-subtraction software 
described by \citet{kelson03}. All images were overscan-subtracted and 
a simple cosmic-ray cleaning algorithm was applied to the individual cluster 
exposures.  The images were also rotated such that the dispersion direction was 
approximately parallel to the rows of the detector. Inspection of the several 
sets of flatfield exposures obtained over the course of the run showed no 
measurable variation and all of the flatfield exposures were therefore summed 
to produce one master flatfield. This flat was then normalized with low-order 
fits to each order and divided into all of the object exposures prior to 
sky subtraction. 

Sky subtraction of MIKE data is complicated by the nonlinear transformation 
between the orthogonal (dispersion, spatial) and ``natural'' detector 
(row, column) coordinate systems and the slight undersampling of the data. 
The marginally undersampled sky lines are tilted with respect to the 
detector coordinate system and this tilt angle varies as a function of 
spatial position. As discussed in detail by \citet{kelson03}, the classical 
approach of rectifying and rebinning the image prior to sky subtraction can 
result in significant residuals in the case where the sky lines are 
critically- or undersampled. Superior sky subtraction can be obtained if a sky 
model is calculated and subtracted prior to any rectification and rebinning 
of the data. 

The key to this technique is to calculate the transformation between 
the array coordinates $(x,y)$ and the orthogonal dispersion and spatial 
coordinate system $(x_r,y_t)$. By collapsing the spectrum along the spatial 
coordinate $y_t$, the profile of each sky line can then be measured as a 
function of the coordinate $x_r$ in which the sky line is well-sampled (due 
to the tilt of the observed spectrum). A two-dimensional sky model 
can then be constructed with the same pixelization as the original data, 
thus avoiding the residuals that result from rebinning marginally or 
undersampled data. To employ Kelson's method, we first calculated 
the transformation $y_t = Y(x,y)$ with his {\tt getrect} command, using a 
flatfield frame to trace the boundaries of each order. These boundaries 
were defined with {\tt findslits} (treating them as individual slits 
in a multislit mask). We then calculated $x_r = X(x,y_t)$ with 
{\tt getrect} applied to each order. Accurate calculation of this 
transformation requires several lines per order. As there are few 
airglow lines in the higher orders (shorter wavelengths), we created a 
composite image of spectroscopic lines by combining a very high SNR sky frame 
(calculated by summing a large number of our exposures of the fainter 
clusters) with ThAr and NeAr lamp spectra. As many ThAr lines saturated in the 
13 reddest orders, these orders were masked out before construction of the 
composite line image for {\tt getrect}. A sky model image for each globular 
cluster exposure was then calculated with {\tt skyrect} and subtracted from 
the flatfielded data. Observations of the template stars were sufficiently 
short that no sky subtraction was deemed necessary. 

\begin{figure*}[!ht]
\epsscale{1.2} 
\plotone{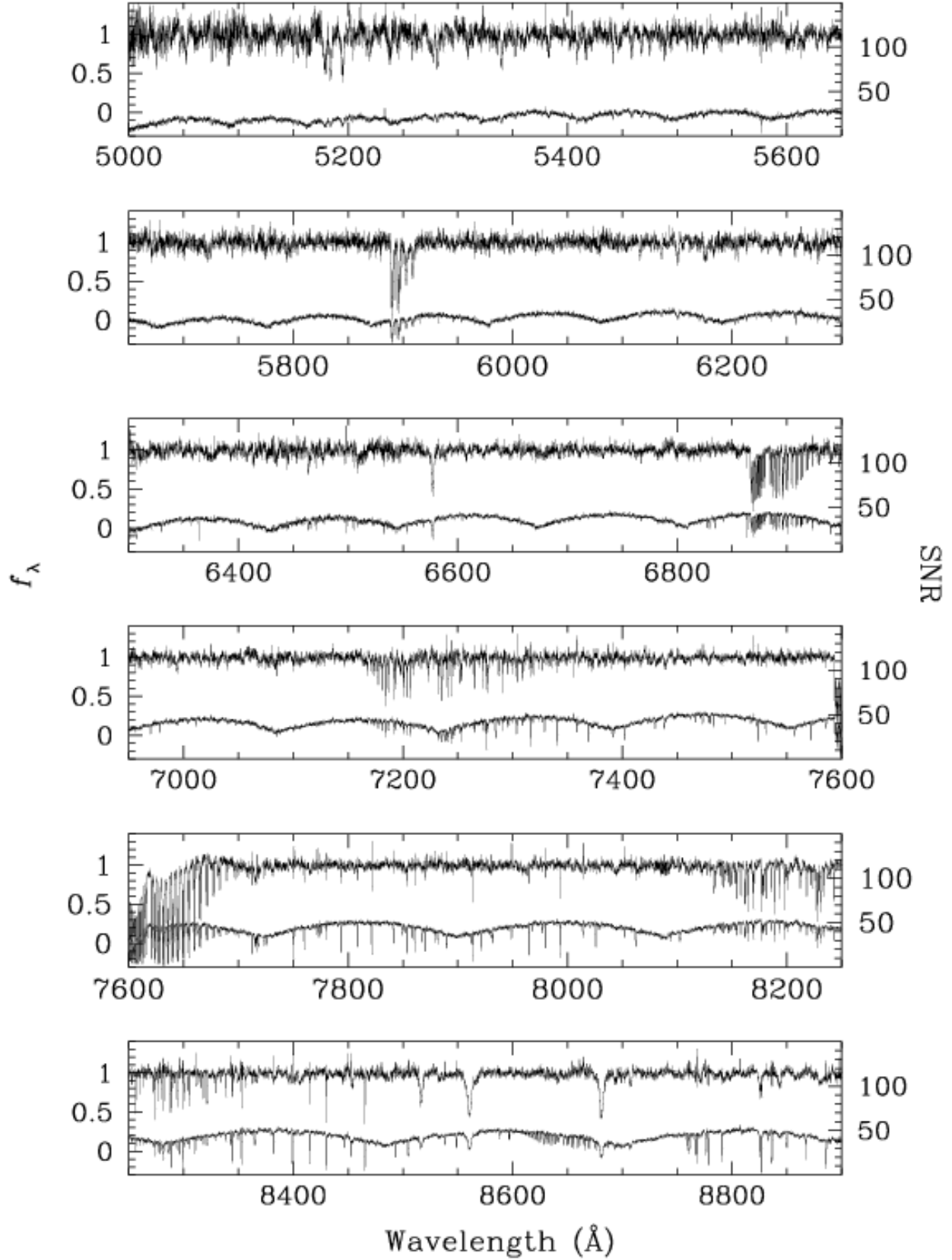} 
\caption{\small 
Orders 39 -- 68 of the \cena\ globular cluster C23. Each order has been 
divided by a low-order polynomial fit to normalize the spectrum ({\it dark 
solid line, left axis}). The SNR as a function of wavelength is shown below 
the spectrum ({\it light solid line, right axis}). The smooth SNR variation 
is due to the transmission of the individual orders, which have been stiched 
together. 
\label{fig:c23}
}
\end{figure*}

\begin{figure*}[!ht]
\epsscale{1.2} 
\plotone{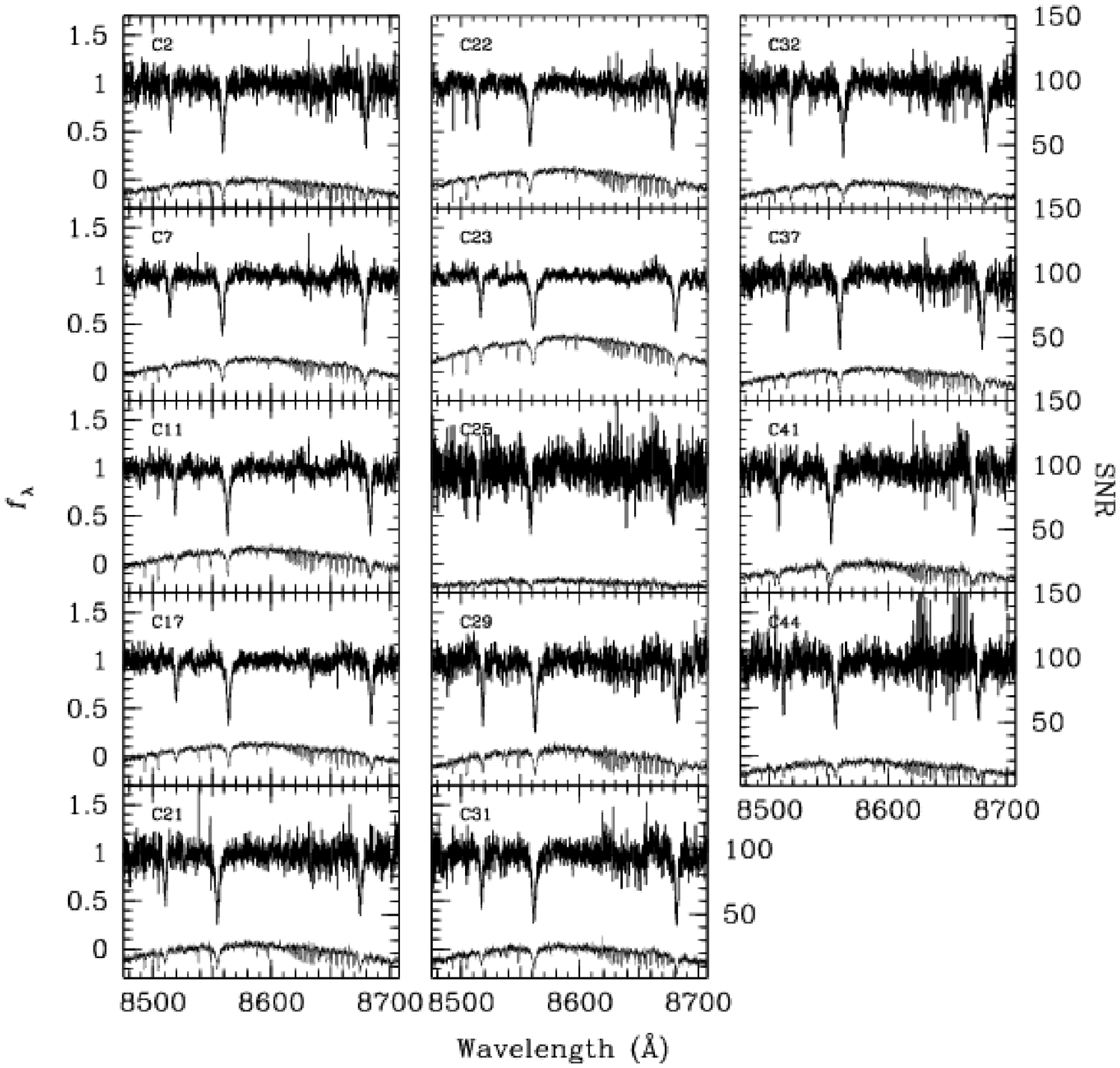} 
\caption{\small 
Calcium triplet region (order~40) for all 14 \cena\ globular clusters. 
\label{fig:cat}
}
\end{figure*}

The dispersion solution was calculated with the {\tt ecidentify} task in the 
IRAF echelle package and ThAr lamp spectra. Although most of the reddest 
orders were saturated by bright lines as described above, a combination of 
multiple lines in the bluer orders and several unsaturated lines in the 
redder orders led to a good solution (rms $\sim$ 0.005 -- 0.007 \AA) with a 
fourth-order Legendre polynomial in $x$ and $y$. 
All of the object spectra were then extracted with {\tt apall}. Because 
most of the individual cluster spectra were too faint for a good trace, 
particularly in the bluest orders, we used a reference trace from a brighter 
object. This trace was recentered based on the centroid of the cluster trace 
in the reddest, highest SNR orders. Each individual exposure was extracted in 
this manner as several-pixel shifts were observed between a small number of 
the multiple exposures on individual clusters. The width of the spectral 
extraction window was 9 pixels or 2.6$''$.  The arc lamp was extracted for 
each object with the same trace and the dispersion solution was calculated 
with the {\tt ecreidentify} task. The spectra were then placed on a linear 
wavelength scale with the {\tt dispcor} task for measurement of their 
line-of-sight velocity dispersion $\sigma$. Template stars and the flux 
standard HR~1544 were extracted in a similar manner. Orders 68 -- 39 of C23, 
the cluster with the highest SNR, are shown in 
Figure~\ref{fig:c23}. 
The calcium triplet region (order 40) is shown for all 14 clusters 
in Figure~\ref{fig:cat}. No correction for telluric absorption was applied to 
these data. 

\section{Velocity Dispersion Measurement}  \label{sec:losvd}

The two most commonly employed techniques for velocity dispersion measurement 
are direct-fitting in pixel space \citep[\eg][]{rix92,kelson00,barth02} and 
cross-correlation techniques in the Fourier domain 
\citep[\eg][]{illingworth76}. 
After experiments with both techniques we chose the cross-correlation method 
developed in detail by \citet{tonry79} as implemented by the {\tt rvsao} 
package \citep{kurtz98} in IRAF. The main advantage of the cross-correlation 
function (hereafter CCF) is that it is less sensitive to the relative line 
strengths in the targets and templates. This is likely to be the case for the 
present study due to metallicity differences between the globular clusters 
and the Galactic template stars observed. Trial implementations of 
direct-fitting in pixel space confirmed that most of the stellar templates 
are not good matches in detail to these globular clusters. Direct 
pixel-fitting did show that the observed stellar templates with spectral types 
between G8 and K3 were the best matches to the globular cluster spectral 
energy distributions. The stellar templates used for this study are: 
HD~80499 (G8III), HD~95272 (K0III), HD~88284 (K0III), HD~43827 (K1III), 
HD~92588 (K1IV), HD~44951 (K3III), and HD~46184 (K3III). 

The first step in implementation of the CCF technique is the calculation of 
the relation between the CCF width and the observed line-of-sight velocity 
dispersion $\sigma$. To measure this relationship, 
we convolved a range of the observed stellar templates by Gaussian functions 
with $\sigma =$ [5,7,10,14,20,28,40,56] \kms and used a simple spline fit to 
calculate the relation between CCF width and input $\sigma$. 
We calculated this relationship for each of the spectroscopic orders and 
determined that the uncertainty in $\sigma$ measurements in orders with 
significant telluric absorption (50, 48, 47, 45, 42, and 41) was considerably 
higher than those without. Given the large number of orders available, these 
orders were excluded from subsequent analysis. 
Based on measurements from the remaining orders we determined 
that all of the globular clusters have $\sigma$ in the approximate range
of 10 -- 30 \kms. The velocity dispersions of these clusters are thus 
well-resolved by this instrument configuration. 

\begin{figure*}
\plotone{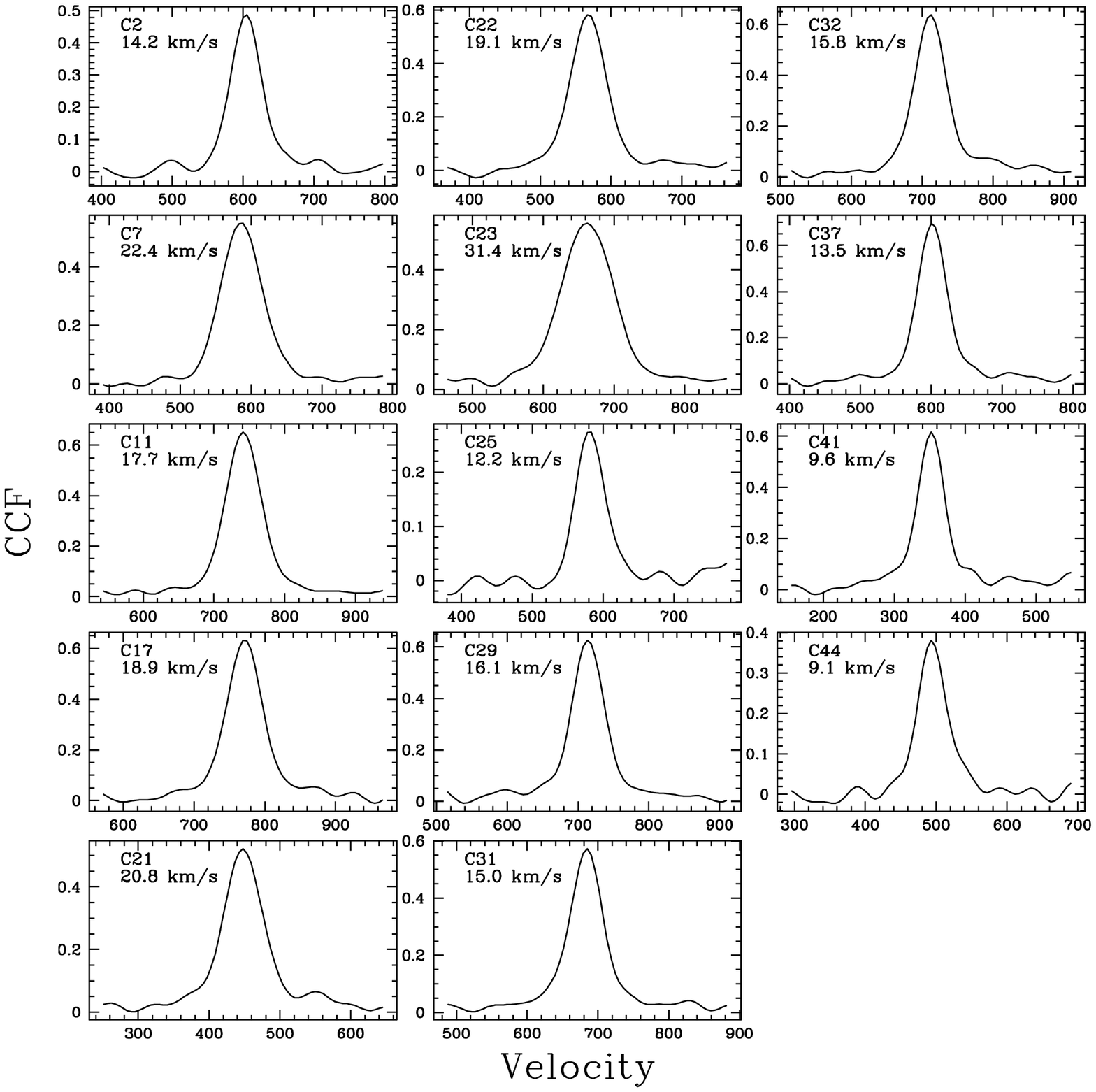} 
\caption{\small 
Cross-correlation function for each globular cluster over 
the wavelength range $5000 - 6800$ \AA\ and binned to 0.2 \AA\ pix$^{-1}$. 
The derived velocity dispersion is listed in the top left corner of each panel. 
\label{fig:ccf}
}
\end{figure*}

While the CCF technique is known to be relatively insensitive to template 
mismatch, we tested the importance of this potential source of systematic 
error by convolving each spectral type by the range of velocity dispersions 
given above 
and using other templates of similar but not identical temperature to measure 
$\sigma$ (\eg\ a K0III template to measure an artificially broadened K3III 
template). This analysis showed that template mismatch produced at most an rms 
uncertainty of 1 \kms\ when $\sigma$ was calculated for a single order. 
Given the large number of orders available and our estimate of the 
range in $\sigma$ of these clusters, we then experimented with combining 
multiple orders together as well as binning the data to improve SNR. 
A range of tests showed we obtain excellent and very stable results 
over the wavelength range 5000 -- 6800 \AA\ and the spectra rebinned to 
a scale of 0.2 \AA\ pixel$^{-1}$. The rms variation due to template mismatch 
over this range was less than 0.5 \kms. CCFs for all 14 clusters are 
shown in Figure~\ref{fig:ccf} and the measured $\sigma$ and estimated 
uncertainties are listed in Table~\ref{tbl:obs}. 

The 5000 -- 6800 \AA\ spectral range does not include the higher SNR CaT 
region. While the CaT lines are quite strong for all of the clusters 
(see Figure~\ref{fig:cat}), these lines are in fact so strong that they 
exhibit significant damping wings. These damping wings broaden the line 
profiles of the template stars significantly and they are not a good 
representation of the instrumental line profile. The CaT lines in the template 
stars have intrinsic widths comparable to or larger than the measured 
velocity dispersions of the globular clusters over the 5000 -- 6800 \AA\ 
region. We therefore did not use these lines to measure the cluster 
velocity dispersions. The neighboring, red orders have comparable SNR to the 
CaT region, although these orders are contaminated by significant telluric 
absorption. Although the telluric absorption is to some extent correctable, 
these orders were also discarded for the velocity dispersion measurement  
because the SNR in the 5000 --- 6800 \AA\ region is sufficiently high that SNR 
is at most a modest contributor to the total error budget (see below). 
While there are still reasonably strong lines in the 5000 -- 6800 \AA\ region, 
notably Mg$b$ and H$\alpha$, these lines only make a minor contribution 
to the velocity dispersion measurement. 

\begin{figure}
\plotone{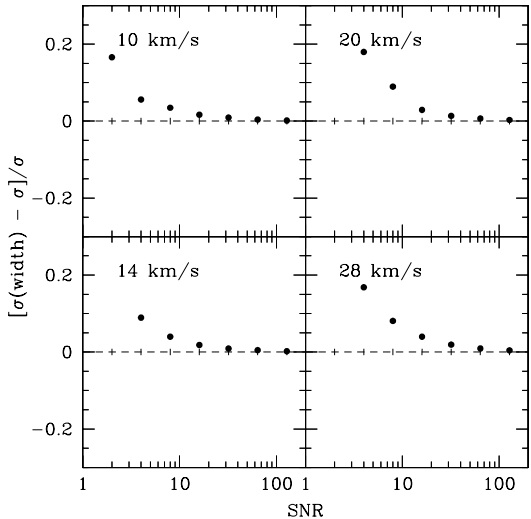} 
\caption{\small 
Velocity dispersion residuals as a function of signal-to-noise ratio 
for the spectral region 5000 -- 6800 \AA\ and binned to 0.2 \AA\ pix$^{-1}$. 
The $\sigma$ values shown closely correspond to the observed range for the 
\cena\ globular clusters. Less than half of the cluster observations have 
mean SNRs below ten. Tickmarks along the dashed line mark the location of 
the model SNR values. Only the panel for $\sigma = 10$ \kms plots a residual 
value for SNR$=2$. The residuals were larger than 30\% at SNR$=2$ for the 
remaining three velocity dispersions. 
\label{fig:snr}
}
\end{figure}

Given the range of SNR and $\sigma$ of these data, we performed simulations 
with our template stars to determine the quality of our measurements 
as a function of both SNR and $\sigma$. These tests were performed by first 
convolving the best-fitting template (HD~88284) with the range of velocity 
dispersions 
described above. We then added noise to these convolved spectra to 
produce a set of output spectra with average SNR of [2,4,8,16,32,64,128] for 
each velocity dispersion. As the SNR per pixel is quite variable due to both 
the blaze 
function and the red color of these clusters, we used an input SNR spectrum 
from one of the cluster observations 
(C23, shown in Figure~\ref{fig:c23}) 
to reproduce the pixel-to-pixel variation in the SNR, while 
forcing the mean SNR in the spectra to equal the specified values. 
The widths of these input spectra were then calculated in a similar fashion 
to the globular cluster measurements described above. The results of 
this analysis for $\sigma = 10 - 28$ \kms (the range most appropriate to 
this study) are shown in Figure~\ref{fig:snr}. This figure displays the 
difference between the input $\sigma$ and the measured $\sigma$ at a given 
SNR, normalized by the input $\sigma$. 

As expected, the deviation between the input $\sigma$ and the measured $\sigma$
increases toward lower SNR. In particular, these simulations show that 
at lower SNR the measurements will tend to overestimate $\sigma$. This 
may be because the addition of noise systematically erases narrower features 
more readily than broader features. The simulations also show that at a given 
SNR the velocity dispersion measurement will be more accurate 
if the intrinsic velocity dispersion is lower. This is likely a reflection of 
the net SNR over the number of pixels that comprise a given feature. The fewer 
pixels that span, \eg\ the absorption lines employed here, the higher the net 
SNR of the feature relative to a spectrum of the same mean SNR per pixel with 
a larger $\sigma$. This systematic effect on the $\sigma$ measurement impacts 
our observations at most at the 5\% level. Three of the clusters with 
velocity dispersions of on order 10 \kms (C25, C41, C44) have a mean SNR less 
than 10 over then 5000 -- 6800 \AA\ range in our rebinned spectra, while three 
(C2, C31, C32) in the 14 \kms range have a mean 
SNR $\sim 10$ (see Table~\ref{tbl:obs}). 
The measured velocity dispersions for these low-SNR clusters have been 
adjusted to account for the expected overestimate.  
The remaining clusters do not suffer from systematic errors due to low SNR. 

The resolution of our rebinned spectra could also impact the measured $\sigma$ 
as the velocity dispersion resolution now varies from 12 \kms at 
5000 \AA\ to 8.8 \kms at 6800 \AA. However, measurements of $\sigma$ with 
a dispersion of 0.1 \AA\ pixel$^{-1}$ yield consistent results with 
the values in Table~\ref{tbl:obs}, although with somewhat larger scatter 
due to the lower SNR. The relationship between measured $\sigma$ and 
SNR at each $\sigma$ shown in Figure~\ref{fig:snr} also demonstrates that 
the measured $\sigma$ converges to the input value as the SNR increases. 
The measured values quoted in Table~\ref{tbl:obs} also agree well with 
measurements obtained from single, unbinned orders and direct-pixel fitting, 
although the latter approaches have larger scatter. 
We note that preliminary velocity dispersions measurements from high-resolution 
spectroscopy of four \cena\ globular clusters were obtained by 
Dubath and reported in \citet{harris02}. 
These measurements are marginally consistent with our own. 

The $1'' \times 2.6''$ spectral extraction window we employed corresponds to 
a physical area of $18.6 \times 48.4$ pc, which is larger than the half-light 
radius for all of these clusters (\eg\ see Table~\ref{tbl:params}).
In King models for globular clusters, the 
velocity dispersion is a function of radius. Our measurements, because they 
include most of the cluster, should correspond well to the global, 
one-dimensional velocity dispersion $\sigma_\infty$. In contrast, the 
core one-dimensional velocity dispersion $\sigma_0$ is approximately the 
quantity measured for Galactic globular clusters and the most commonly quoted 
quantity in the literature. The core value $\sigma_0$ is larger than 
$\sigma_\infty$ in the King 
models and the degree of difference depends on cluster concentration. 
For the clusters studied here, this correction is on order 5 -- 10\% for the 
range of cluster concentrations in this sample \citep[see \eg][]{djorgovski97}. 
The errorbars quoted in Table~\ref{tbl:obs} include the uncertainties in this 
correction, which are caused by uncertainties in the measured structural 
parameters reported by \citet{harris02} and the exact placement of the 
slit, as well as the uncertainties due to template mismatch and SNR. 

\section{Analysis} 

\subsection{Observed Cluster Properties} 

These 14 globular clusters are the brightest clusters with King model 
structural parameters measured with \hst\ by \citet{harris02}. 
These models are 
described by the parameters $W_0$ (central potential), $r_c$ (core radius) 
and $c$ (cluster concentration, where $c = {\rm log}\,r_t/r_c$ and $r_t$ is 
the tidal radius). \citet{harris02} also fit for ellipticity. 
Both $r_c$ and $r_h$ (the model half-mass radius) 
from their fits are provided in Table~\ref{tbl:params}, converted to 
parsecs, and their measurement of $c$ is also given. 
The projected half-light radius is 
$R_h \approx 0.73 \, r_h$. As these values span the approximate 
range of $1.4 < c < 2.1$, none are core-collapse clusters. 
All 14 of these clusters were also included in the recent study by 
\citet{peng04a}, who have obtained $UBVRI$ photometry and radial velocities 
for over 200 \cena\ globular clusters. As most of the \citet{harris02} 
photometry is from unfiltered STIS images, we have listed instead 
the \citet{peng04a} $V_0$, $(B-V)_0$, and $(V-I)_0$ values in the table, where 
these 
quantities have been dereddened based on their quoted $E(B-V)$ and assuming 
$R_V = 3.1$. We also include the central surface brightness $\mu_V^0$ from 
\citet{harris02}, but recalibrated by the difference between their quoted 
$V$ magnitudes and those in \citet{peng04a}, and have calculated the 
mean, reddening-corrected surface brightness within the projected half-light 
radius: $\langle\mu_V\rangle_h = V_0 + 0.75 - 2.5 \, \log \, \pi R_h^2$. All of these 
dereddened values only account for Galactic dust and do not include any dust 
intrinsic to \cena. While most of these clusters are not near \cena's 
prominent dust lane, dust may affect the colors of some of these clusters. 

\begin{center}
\begin{deluxetable*}{lccccccccc}
%\tabletypesize{\scriptsize} 
\tablecolumns{10}
\tablewidth{0pt}
\tablecaption{Structural and Photometric Measurements\label{tbl:params}}
\tablehead{
\colhead{ID} &
\colhead{$r_c$ [pc]} &
\colhead{$r_h$ [pc]} &
\colhead{$c$} &
\colhead{$\mu_V^0$} & 
\colhead{$\langle\mu_V\rangle_h$} & 
\colhead{$V_0$} &
\colhead{$M_V$} &
\colhead{$(B-V)_0$} & 
\colhead{$(V-I)_0$} \\
}
\startdata
C2      . . . . . . . . . . & 0.80 & 8.49 & 1.99 & 16.27 & 17.70 & 18.10 & -9.82 & 0.70 & 0.86   \\
C7$^x$  . . . . . . . . . . & 1.41 & 10.00 & 1.83 & 15.46 & 16.79 & 16.83 & -11.09 & 0.75 & 0.91 \\
C11     . . . . . . . . . . & 1.30 & 10.41 & 1.88 & 16.19 & 17.58 & 17.54 & -10.38 & 0.93 & 1.11 \\
C17     . . . . . . . . . . & 2.27 & 7.60 & 1.43 & 16.14 & 16.65 & 17.29 & -10.63 & 0.77 & 0.88  \\
C21     . . . . . . . . . . & 1.21 & 9.27 & 1.86 & 15.97 & 17.32 & 17.53 & -10.39 & 0.78 & 0.93  \\
C22     . . . . . . . . . . & 1.10 & 5.06 & 1.62 & 15.60 & 16.29 & 17.81 & -10.11 & 0.79 & 0.91  \\
C23$^x$ . . . . . . . . . . & 0.87 & 4.41 & 1.67 & 14.37 & 15.06 & 16.88 & -11.04 & 0.96 & 1.10  \\
C25$^x$ . . . . . . . . . . & 0.99 & 8.32 & 1.90 & 16.34 & 17.71 & 18.15 & -9.77 & 0.96 & 1.14   \\
C29$^x$ . . . . . . . . . . & 1.19 & 9.16 & 1.87 & 16.21 & 17.52 & 17.75 & -10.17 & 0.87 & 1.05  \\
C31     . . . . . . . . . . & 0.91 & 5.29 & 1.74 & 15.71 & 16.58 & 18.01 & -9.91 & 0.92 & 1.12   \\
C32$^x$ . . . . . . . . . . & 0.56 & 7.24 & 2.06 & 15.88 & 17.33 & 18.07 & -9.85 & 0.94 & 1.10   \\
C37$^x$ . . . . . . . . . . & 0.58 & 4.43 & 1.87 & 15.31 & 16.28 & 18.09 & -9.83 & 0.84 & 0.99   \\
C41     . . . . . . . . . . & 0.78 & 5.99 & 1.87 & 15.94 & 17.04 & 18.19 & -9.73 & 0.87 & 1.06   \\
C44     . . . . . . . . . . & 1.25 & 7.60 & 1.70 & 16.45 & 17.68 & 18.32 & -9.60 & 0.68 & 0.84   \\
\enddata
\tablecomments{
Structural and photometric parameters of our sample based on measurements 
from the literature. 
For each cluster in column~1 we list $r_c$, $r_h$, $c$, and $\mu_V^0$ 
from \citet{harris02} in cols.~2--5, where we have adopted a distance to 
\cena\ of 3.84~Mpc. Col.~6 contains the mean surface brightness within the 
projected half-light radius, while cols~7--10 are the reddening-corrected $V_0$ 
magnitude, the absolute $V$ magnitude, and the $(B-V)_0$ and $(V-I)_0$ colors 
from \citet{peng04a}. 
We have used this photometry to apply a correction to $\mu_V^0$ (see text) 
and calculate $\langle\mu_V\rangle_h$. The six clusters marked with an $x$ 
superscript indicate that \citet{harris02} found evidence for extratidal light 
in their surface photometry. 
}
\end{deluxetable*}
\end{center}

\begin{figure}
\plotone{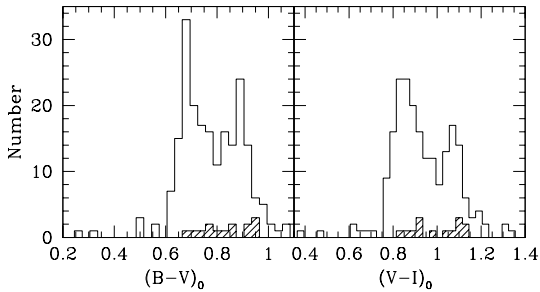} 
\caption{\small 
Reddening-corrected colors of \cena\ globular clusters. 
All clusters observed by \citet{peng04a} are shown in the histogram, 
while the globular clusters in the present study are in the hatched histogram. 
The latter clusters span a range in color representative of the \cena\ globular 
cluster system. 
\label{fig:colors}
}
\end{figure}

\begin{figure*}[!ht]
\plotone{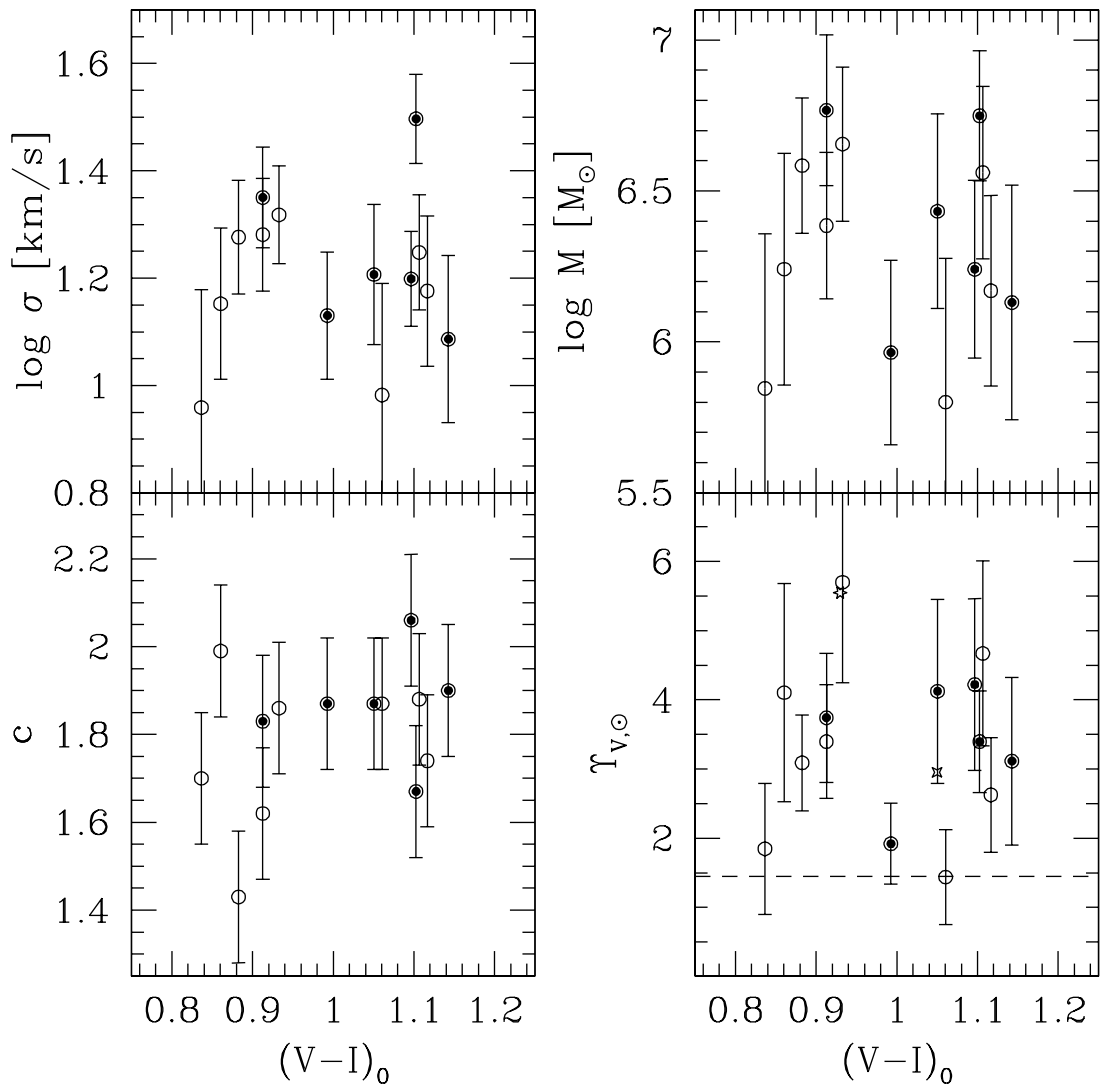} 
%\epsscale{1.2}
\caption{\small 
Reddening-corrected $V-I$ color of \cena\ globular clusters as a function of 
velocity dispersion ({\it upper left}), central concentration 
({\it lower left}), virial mass ({\it upper right}), and mass-to-light ratio 
({\it lower right}). All \cena\ clusters are marked with circles, while 
clusters identified by \citet{harris02} to possibly have extratidal light 
are marked with partly filled circles. 
The dashed line in the lower-left panel represents the mean core mass-to-light 
ratio derived by \citet{mclaughlin00} for Galactic globular clusters. 
The 0.18 magnitude uncertainty in the \cena\ distance modulus 
\citep{rejkuba04} translates into an additional, systematic uncertainty of 
$\sim 18$\% in the mass-to-light ratio. $\omega$Cen ({\it cross}) and 
G1 ({\it star}) are also shown in this panel. 
\label{fig:vscolors}
}
\end{figure*}

The color distribution of \cena\ globular clusters is bimodal 
\citep{held97,peng04a}, similar to the globular cluster systems of many 
galaxies \citep[\eg][]{larsen01a,kundu01}. 
A bimodal color distribution is most commonly ascribed to a bimodal 
distribution in metallicity as color is a reasonable proxy for metallicity 
in old single stellar populations. 
In the case of \cena, this galaxy appears to 
have undergone a major merger in the recent past. Based on the strength of 
H$\beta$, \citet{peng04a} conclude that the more metal-rich globular clusters 
formed $5^{+3}_{-2}$ Gyr ago, while \citet{kaviraj04} derive an age estimate 
of 1 -- 2 Gyr for the metal-rich population in their study of the metallicity 
distribution function (derived from the \citet{peng04a} $U$ and $B$ 
photometry). 
Figure~\ref{fig:colors} shows the $(B-V)_0$ and $(V-I)_0$ color distributions 
for all of the globular clusters studied by \citet{peng04a}. The clusters 
in the present study are represented by the hatched histogram and span the full 
range in color of the majority of the \cena\ globular cluster system. 
Figure~\ref{fig:vscolors} demonstrates that these clusters do not show 
any strong correlations between color and velocity dispersion, central 
concentration, mass, or mass-to-light ratio (see next section). 

\subsection{Cluster Masses and Mass-to-Light Ratios}

One of the most valuable and readily calculated properties of a globular 
cluster is its total mass. There are multiple ways to estimate the mass 
of a globular cluster, although the most straightforward approach is to use 
the virial theorem: 
\begin{equation}
M_{vir} = 2.5 \, \frac{\langle v^2\rangle\,r_h}{G}, 
\end{equation} 
where $\langle v^2\rangle \approx 3\sigma^2$ is the mean-square speed 
of the cluster stars and $r_h$ is the half-mass radius \citep{binney87}. 
The virial masses for these clusters (in units of $10^6$ \msun) are listed in 
Table~\ref{tbl:mass}. 

\begin{center}
\begin{deluxetable}{lccccc}
\tablecolumns{6}
\tablecaption{Masses and Mass-to-Light Ratios\label{tbl:mass}}
\tablehead{
\colhead{ID} &
\colhead{$M_{vir}$ [M$_{\odot,6}$]} &
\colhead{$M_{King}$ [M$_{\odot,6}$] } &
\colhead{$\mu_{King}$} &
\colhead{$\Upsilon_{V,\odot}$} &
\colhead{log $E_b$ [erg]} \\
}
\startdata
C2       & $ 2.99 \pm 0.67 $ & 1.74 & 52.4 & $ 4.1 \pm 1.6 $ & 52.16 \\
C7$^x$   & $ 8.75 \pm 1.46 $ & 5.85 & 39.7 & $ 3.7 \pm 0.9 $ & 53.10 \\
C11      & $ 5.69 \pm 1.04 $ & 3.64 & 43.0 & $ 4.7 \pm 1.3 $ & 52.68 \\
C17      & $ 4.73 \pm 0.86 $ & 3.84 & 21.7 & $ 3.1 \pm 0.7 $ & 52.82 \\
C21      & $ 7.00 \pm 1.15 $ & 4.52 & 41.7 & $ 5.7 \pm 1.5 $ & 52.92 \\
C22      & $ 3.22 \pm 0.59 $ & 2.43 & 28.6 & $ 3.4 \pm 0.8 $ & 52.61 \\
C23$^x$  & $ 7.59 \pm 1.21 $ & 5.61 & 30.9 & $ 3.4 \pm 0.7 $ & 53.40 \\
C25$^x$  & $ 2.16 \pm 0.52 $ & 1.35 & 44.5 & $ 3.1 \pm 1.2 $ & 51.93 \\
C29$^x$  & $ 4.14 \pm 0.87 $ & 2.71 & 42.3 & $ 4.1 \pm 1.3 $ & 52.47 \\
C31      & $ 2.07 \pm 0.47 $ & 1.48 & 34.4 & $ 2.6 \pm 0.8 $ & 52.16 \\
C32$^x$  & $ 3.15 \pm 0.51 $ & 1.74 & 60.8 & $ 4.2 \pm 1.2 $ & 52.24 \\
C37$^x$  & $ 1.41 \pm 0.28 $ & 0.92 & 42.3 & $ 1.9 \pm 0.6 $ & 51.85 \\
C41      & $ 0.96 \pm 0.30 $ & 0.63 & 42.3 & $ 1.4 \pm 0.7 $ & 51.39 \\
C44      & $ 1.10 \pm 0.36 $ & 0.70 & 32.3 & $ 1.8 \pm 0.9 $ & 51.41 \\
\enddata
\tablecomments{
Derived parameters for \cena\ globular clusters. For each cluster in 
column~1, the virial mass and King model masses (in units of $10^6$ \msun) 
are listed in columns 2 and 3. The dimensionless King mass, mass-to-light 
ratio, and binding energy are provided in columns 4 -- 6. 
}
\end{deluxetable}
\end{center}

An alternative approach is to use the King parameters to calculate the cluster 
mass. From \citet{illingworth76}, the mass is
\begin{equation}
M_{King} = 167\,r_c\,\mu\,\sigma_0^2, 
\end{equation} 
where $\mu$ is the dimensionless King model mass \citep{king66} and 
$\sigma_0$ is the core velocity dispersion \citep[see also][]{richstone86}. 
As discussed previously, we have estimated $\sigma_0$ from the cluster 
concentration parameter \citep{djorgovski97}. 
The dimensionless mass $\mu$ is also a function of concentration. 
The King model masses for these clusters are $\sim 50$\% less than 
the virial mass estimates. It is not unusual for these two estimates to 
disagree \citep[\eg\ see][for $\omega$Cen and G1]{meylan01}. 
The uncertainties (not shown) in the King masses are larger than the virial 
mass estimates due to uncertainties in the measurement of $c$ and $r_c$, in 
addition to the uncertainties in $\sigma$. We therefore use the virial mass 
in the subsequent discussion. 

Once the mass has been estimated, the $V-$band mass-to-light ratio 
in solar units $\Upsilon_{V,\odot}$ is readily calculated. The virial 
mass is divided by the $V-$band luminosity $L_{V,\odot}$, which is computed 
from the reddening-corrected magnitude $V_0$ (assuming $M_{V,\odot} = 4.83$). 
The mass-to-light ratios for these clusters are listed in column~5 of 
Table~\ref{tbl:mass}. 
These values fall in the range 1.4 -- 5.7 and have an 
average value of $\langle\Upsilon_{V,\odot}\rangle \sim 3$. This average is 
larger than the mean core mass-to-light ratio of $1.45 \pm 0.1$ for globular 
clusters in the Galaxy \citep{mclaughlin00} and the comparable core and 
global values of $1.53 \pm 0.18$ from four globular clusters in 
M33 \citep{larsen02}. 
If we had adopted the King mass estimates to compute the 
mass-to-light ratios, rather than the virial mass estimates, the mass-to-light 
ratios would be $\sim 50$\% less. 
The difference between the \cena\ globular clusters and those in the Local 
Group may be due to either intrinsic differences or due to measurement 
uncertainties. 
If the latter is the case, then the masses are 
overestimated, the luminosities are underestimated, or both. An overestimate 
in the mass could be explained by either an overestimate of $\sigma$ by 
$\sim$ 50\% or an overestimate of $r_h$ by a factor of 2, although both 
of these possibilities vastly exceed the estimated uncertainties in these 
quantities. The mass-to-light ratios between the different globular cluster 
systems could also be brought into agreement if the luminosities are 
underestimated by a factor of 2. This exceeds the estimated uncertainty in 
the distance to \cena\ \citep{rejkuba04}, although intrinsic reddening could 
also contribute and \cena\ is well-known for its prominent, large dust lanes. 
The luminosities could also be underestimated if there is a 
significant contribution to the surface brightness at large radii, where the 
SNR is poor. While a conspiracy of all of these potential sources of 
error could bring the mean mass-to-light ratios of these globular cluster 
systems into agreement, we conclude that the differences are likely real. 

Further support of these large mass-to-light ratios is provided by some of 
the most massive 
globular clusters in the Local Group, including the Galactic globular 
cluster $\omega$Cen, which has $\Upsilon_{V,\odot} = 2.4 - 3.5$ 
and G1 in M31 with $\Upsilon = 3.6 - 7.5$ \citep{meylan95,meylan01}. 
The lower-right panel of Figure~\ref{fig:vscolors} shows the mass-to-light 
ratios for the \cena\ globular clusters as a function of $(V-I)_0$. No obvious 
trend with color is apparent, as might be expected if the generally redder, 
more metal-rich clusters identified by \citet*{peng04b} formed more recently.  
The massive, Local Group clusters $\omega$Cen and G1 are also shown for 
comparison. For these clusters the average of the virial and King mass 
estimates from \citet{meylan01} were used for the mass-to-light ratio, 
while the integrated colors are either from the \citet{harris96} compilation 
($\omega$Cen) or from \citet{heasley88}. 

\subsection{Fundamental Plane}

\cena\ has a significantly larger population of the most massive and luminous 
globular clusters than any galaxy in the Local Group. Our cluster 
sample is dominated by these clusters due to observational constraints, 
yet this population is arguably the most interesting as they provide an 
opportunity to verify and extend relationships for local clusters to more 
extreme examples of the population. The clusters in our sample are 
comparable in mass to $\omega$Cen, the most massive Galactic globular cluster 
at $M = 5 \times 10^6$ \msun \citep{meylan95} and M31's G1, which with 
$M = (7 - 17) \times 10^6$ \msun \citep{meylan01} is the most massive cluster 
known in the Local Group. 

The main relationship of interest is the globular cluster fundamental plane, 
the approximately two-dimensional structure occupied by clusters in the 
three-dimensional space defined by, \eg\ central velocity dispersion, 
surface brightness, and core radius. \citet{djorgovski95} demonstrated that 
the plane occupied by globular clusters is consistent with the expectations 
for virialized cluster cores,  
\begin{equation}
r_c \sim \sigma_0^2 \, I^{-1}_0 \, \Upsilon^{-1}, 
\end{equation} 
and a constant mass-to-light ratio. This appears to be the case for both 
the core properties as well as properties derived at the half-light radius, 
with the differences between these two regimes due to the degree of 
central concentration. 

\begin{figure*}[!ht] 
\plotone{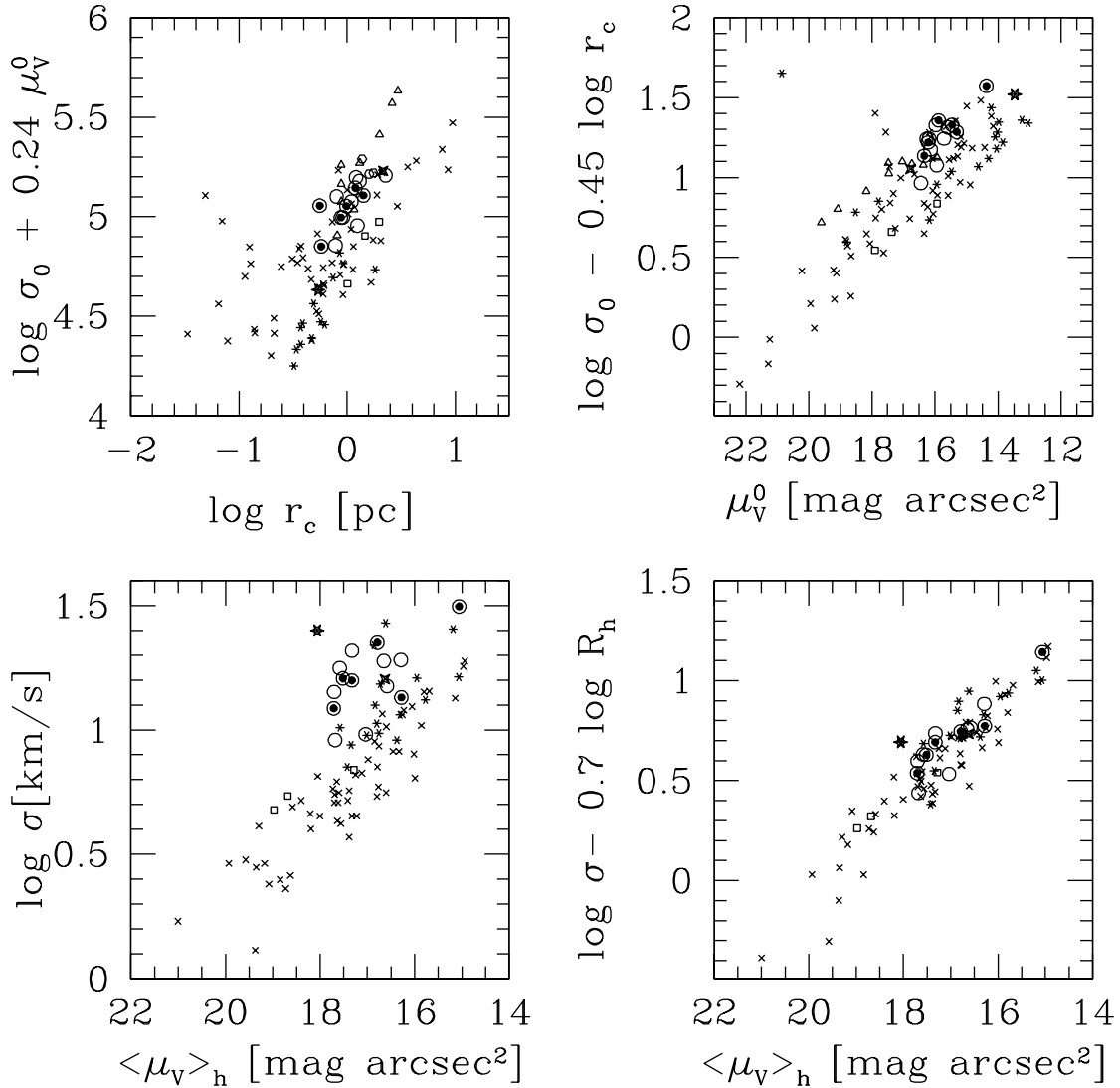} 
\caption{\small 
Core and half-light fundamental plane correlations for globular clusters. 
The core fundamental plane projections ({\it top panels}) demonstrate that the 
\cena\ clusters ({\it open circles}), including those with possible 
extratidal light ({\it partly filled circles}), approximately follow the same 
relations as globular clusters in the Local Group. 
These globular cluster systems include those for our Galaxy ({\it crosses}), 
M31 ({\it stars}), M33 ({\it squares}), the Magellanic Clouds 
({\it triangles}), and the Fornax dwarf galaxy ({\it pentagons}). 
Half-light parameters ({\it bottom panels}) show comparable relations. 
Data were culled from the literature sources cited in the text. 
\label{fig:fp}
}
\end{figure*}

In the top panels of Figure~\ref{fig:fp} we plot two projections of the 
core fundamental plane from \citet{djorgovski95} for the \cena\ globular 
clusters, along with clusters in the Milky Way, M31, M33, and the 
Magellanic Clouds. These projections show that the \cena\ clusters are 
relatively large and have higher surface brightnesses than Galactic clusters, 
although are more similar to globular clusters studied in other Local Group 
galaxies. Half-light projections ({\it bottom panels}) show all of these 
globular cluster systems follow similar trends to the Milky Way system, 
although extragalactic globular clusters appear to have systematically 
lower mean surface brightness within the half-light radius for a given 
$\sigma$. 
This may reflect a bias toward selection of bright objects that nevertheless 
appear marginally resolved in ground-based images or a relative 
overestimate of $R_h$ for more distant clusters. 

Numerous literature sources were employed to obtain data for globular 
clusters in the Local Group. Data for the Milky Way were obtained from 
\citet{pryor93} and \citet{harris96}. Velocity dispersion data for M31 were 
obtained from \citet{dubath97a} and \citet{djorgovski97}, while structural 
parameters were obtained from \citet{barmby02} or earlier work 
\citep{battistini82,crampton85,bendinelli93,fusipecci94,rich96,grillmair96}. 
Data for the Magellanic Clouds were obtained from \cite{dubath97b} and 
data for M33 were obtained from \citet{larsen02}, while the photometry 
for the latter was dereddened with the $E(V-I)$ values derived by 
\citet{sarajedini98}. 
Data for the three Fornax dwarf galaxy's globular clusters with velocity 
dispersion 
measurements and structural parameters were obtained from \citet*{dubath92}
and \citet{mackey03}.  

\begin{figure}[!h] 
\plotone{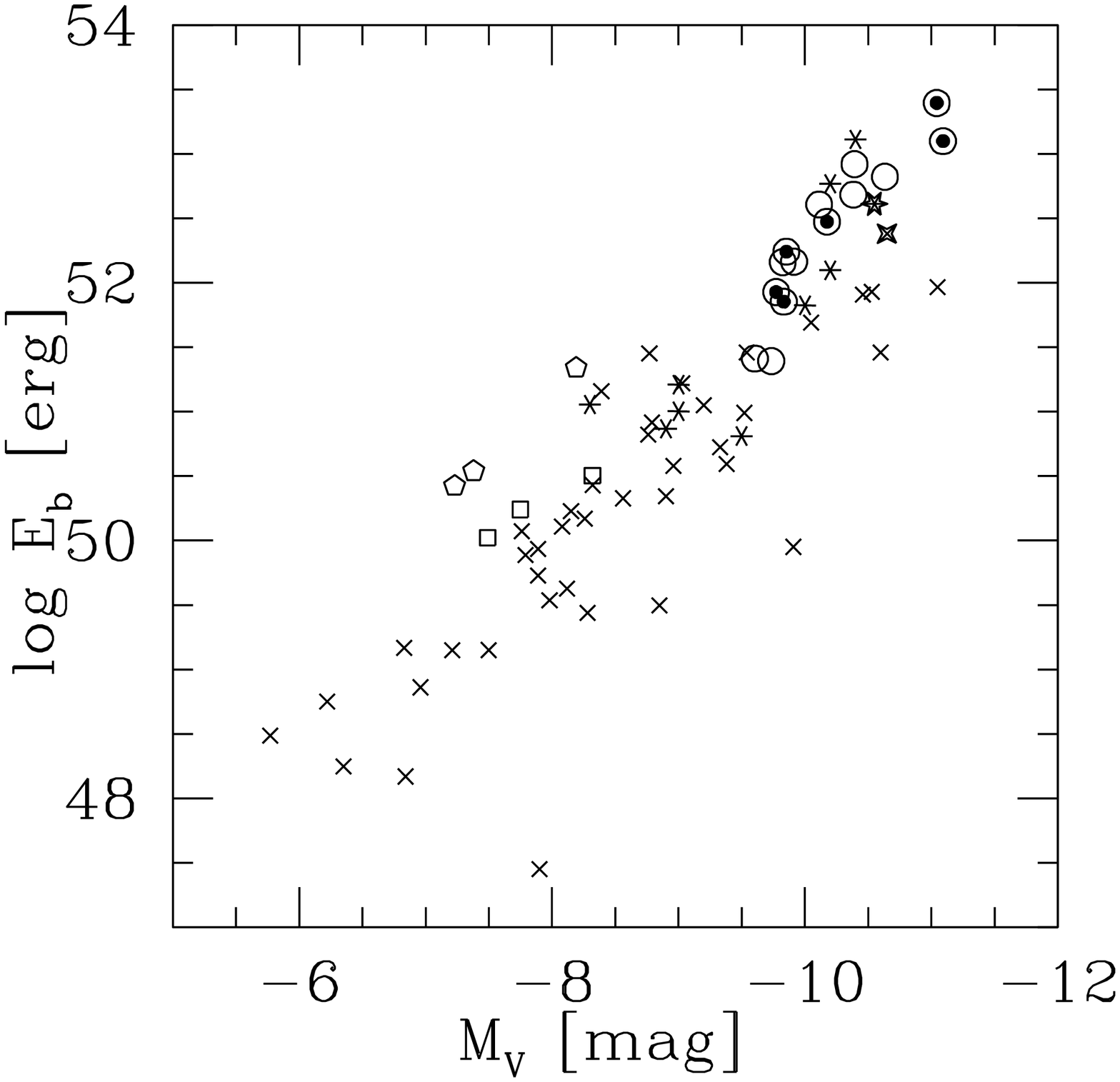} 
\caption{\small 
Binding energy vs. absolute $V$ magnitude for globular clusters. 
The \cena\ globular clusters ({\it circles}) follow approximately the same 
relation found by \citet{mclaughlin00} for Galactic clusters ({\it crosses}), 
although they diverge somewhat toward higher binding energy for a given $M_V$. 
Clusters from M31 ({\it stars}), M33 ({\it squares}), and the Fornax dwarf
galaxy ({\it pentagons}) are also shown. 
$\omega$Cen and M31~G1 are shown as an emphasized cross and star, respectively. 
\cena\ clusters with possible extratidal light are marked with partly 
filled circles. 
\label{fig:eb}
}
\end{figure}

\citet{mclaughlin00} has recently recast discussion of the fundamental 
plane for globular clusters in terms of mass-to-light ratio, luminosity, 
binding energy, and Galactocentric radius. In the context of the four 
independent parameters that characterize single-mass, isotropic King models, 
he found that Galactic globular clusters could be described with a constant 
mass-to-light ratio and a binding energy regulated only by luminosity and 
Galactocentric radius: $E_b = 7.2 \times 10^{39} (L/L_\odot)^{2.05} (r_{gc}/8\,{\rm kpc})^{-0.4}$ erg. Figure~\ref{fig:eb} shows that \cena\ clusters 
follow approximately the same relation between binding energy and luminosity 
as Milky Way and other globular clusters in the Local Group, although 
the \cena\ and other clusters with higher estimated $M/L$ fall above the 
relation traced by Milky Way clusters. 
We did not explore a dependence of binding energy on projected galactocentric 
radius due to distance uncertainties.

\section{Connection to other Spheroidal Systems} 

Many efforts over the years have investigated the connection between globular 
clusters and other spheroidal systems. One valuable approach is  
the $\kappa-$space formalism developed by \citet{bender92}. This space is 
comprised of three orthogonal axes that are proportional to mass ($\kappa_1$), 
mass-to-light ratio ($\kappa_2$), and a third perpendicular axis that scales 
as surface brightness cubed times mass-to-light ratio ($\kappa_3$). 
\citet{burstein97} compiled $\kappa$--space coordinates for a large number of 
spheroidal systems, with masses from globular cluster scales to clusters of 
galaxies. For globular clusters, the $\kappa-$space coordinates are derived as 
\begin{equation}
\kappa_1 = (\log \sigma_e^2 + \log r_e)/\sqrt{2} + 0.11 \\
\end{equation}
\begin{equation}
\kappa_2 = (\log \sigma_e^2 + 2 \, \log I_e - \log r_e)/\sqrt{6} + 0.06 \\
\end{equation}
and 
\begin{equation}
\kappa_3 = (\log \sigma_e^2 - \log I_e - \log r_e)/\sqrt{3} + 0.09, \\
\end{equation}
where $\sigma_e \approx \sigma_\infty$, $r_e \approx r_h$, and 
$I_e$ is the mean surface brightness within $r_e$ in units of $B-$band 
solar luminosities per square parsec. 

\begin{figure*}[!ht] 
\epsscale{1.2} 
\plotone{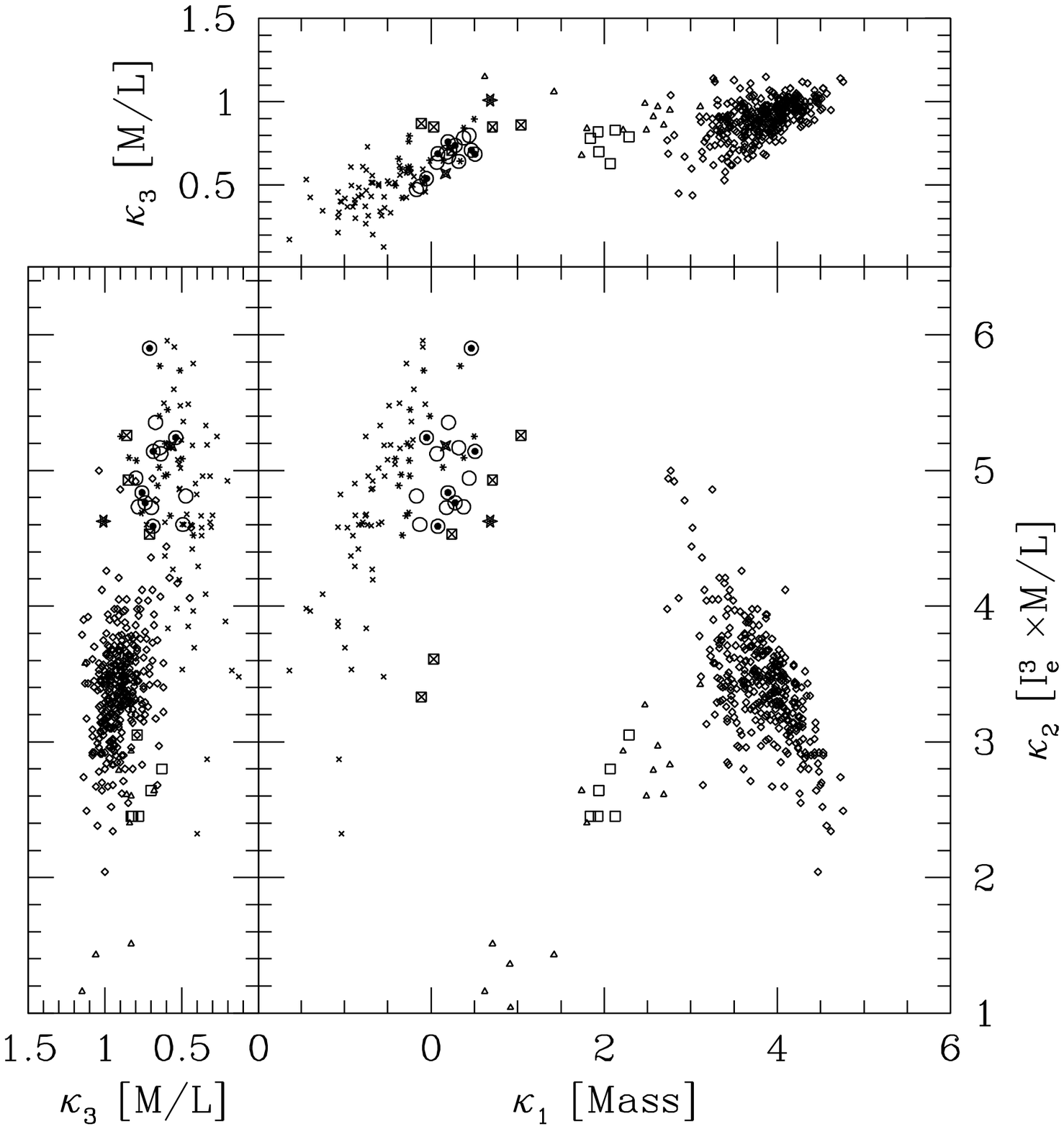} 
\caption{\small 
Projections of 3-dimensional $\kappa-$space. The \cena\ globular clusters 
({\it open, large circles}, or {\it partly filled circles} for those with 
possible extratidal light) are in the region of $\kappa-$space that overlaps 
the most massive Galactic globular clusters ({\it crosses}), M31 clusters 
({\it stars}) and dE nuclei ({\it squares with crosses}). 
$\omega$Cen and M31~G1 are shown as an emphasized cross and star, respectively. 
The integrated properties of the dEs ({\it open squares}) place
them among other dEs and dSphs ({\it open triangles}). Elliptical galaxies and 
bulges are also shown ({\it diamonds}). 
\label{fig:kappa}
}
\end{figure*}

In Figure~\ref{fig:kappa} we plot the three $\kappa-$space projections of the 
fundamental plane for \cena\ clusters ({\it open circles}), along with 
Galactic globular clusters ({\it crosses}), those in M31 ({\it stars}), and 
galaxies \citep[{\it diamonds} and {\it open triangles}, 
data from][]{burstein97}. 
This figure indicates that the \cena\ clusters are quite similar, although 
systematically more massive, than the Galactic clusters. 
In the $\kappa_1$ coordinate, globular clusters are clearly separated 
from galaxies. This does not 
necessarily rule out the existence of objects in this gap, as a 
simple extrapolation of the mass functions for both globular cluster systems 
and dwarf galaxies suggest that such objects should be extremely rare and, 
in the case of dwarf galaxies, quite difficult to detect. Simple scaling 
arguments suggest that globular cluster systems in massive ellipticals such as 
\cena\ are arguably the best place to identify objects in this region,  
provided there is not some other physical mechanism that sets a maximum 
globular cluster mass. 
The globular clusters in the present study are a step toward eliminating the 
mass gap. 

There is also a separation in the $\kappa_3$ coordinate ($M/L$) between 
globular clusters and galaxies, although not as stark as $\kappa_1$ 
(see Figure~\ref{fig:kappa}, {\it left panel}). The mean mass-to-light 
ratio of Galactic globulars is 1.45 \citep{mclaughlin00}, and 
\citet{larsen02} found a comparable number for M33. While the higher 
mass-to-light ratio for \cena\ clusters is somewhat surprising, this may be 
a common feature of more massive clusters. For example, the mass-to-light 
ratios of $\omega$Cen and G1 are in the same range as those in \cena, as 
are the mass-to-light ratios (or $\kappa_3$ values) of other massive 
clusters in the Galaxy and M31. 
The $\kappa_3$ values for dwarf and regular elliptical galaxies are higher 
than for most globular clusters, reflecting their larger mass-to-light ratios, 
although they overlap with the most massive globular clusters. 
The fundamental plane relationship for spheroidal galaxies requires 
$\Upsilon_V \propto L_V^{0.2}$ \citep{vandermarel91,magorrian98}. 
A relationship between mass (or luminosity) and mass-to-light ratio for 
the most massive globular clusters is suggested by their apparently higher 
$\kappa_3$ values. 

The $\kappa_2$ ($I_e^3 \times M/L$) coordinate is essentially an indicator of 
surface brightness at the effective radius as its dependence on $M/L$ is 
much weaker than $I_e$. 
$\kappa_2$ exhibits a great deal of overlap in the properties of all 
spheroidal systems. In fact, the galaxies overlap much more closely with 
Galactic globular clusters 
than those in \cena. This is because $\kappa_2$ is dominated by the mean 
surface brightness within the effective radius and the \cena\ clusters 
are on average lower in mean surface brightness within this radius 
(\eg\ see Figure~\ref{fig:fp}). 

While the \cena\ globular clusters overlap more with other globular 
clusters than galaxies, they are an even better match to the nuclei of the 
dwarf ellipticals studied by \citet{geha02}, who 
measured both integrated and nuclear velocity dispersions and surface 
brightnesses for a sample of Virgo dwarf ellipticals. 
While the integrated 
properties of nucleated dwarf galaxies ({\it large, open squares}) fall within 
the region of $\kappa-$space inhabited by other dwarf ellipticals and 
spheroids ({\it diamonds}), their nuclei ({\it squares with crosses}) overlap 
very well with the \cena\ clusters. 

The $\kappa_1 - \kappa_2$ relation for giant elliptical galaxies and bulges 
exhibits an anticorrelation between mass and surface brightness that sets them 
nearly orthogonal to the dwarf galaxies, which are correlated in $\kappa_1 
- \kappa_2$. This correlation reflects the correlation between luminosity 
and larger cores of higher surface brightness found by \citet{kormendy85}. 
While this correlation suggests mass loss due to winds, \citet{bender92} 
found that mass loss would produce too steep a relation between 
$\kappa_1$ and $\kappa_2$, instead suggesting the observed dwarfs 
are the remnants of a larger, unobserved population. The massive globular 
clusters appear to exhibit a similar $\kappa_1 - \kappa_2$ correlation, 
although steeper than the dwarf galaxies. 

\section{Discussion} \label{sec:discuss}

The strong dynamical and photometric similarities of these massive globular 
clusters to the nuclei of dwarf galaxies has interesting implications for 
models which posit stripped dwarf galaxy nuclei as the origin of 
the most massive globular clusters. The most massive Local Group clusters, 
such as $\omega$Cen in our own Galaxy and G1 in M31, have both been 
discussed and modeled as stripped dwarf nuclei \citep{freeman93,meylan01,
gnedin02,bekki03b}. The position of $\omega$Cen and G1 in Figures~\ref{fig:fp} 
and \ref{fig:kappa} show that the \cena\ clusters occupy a similar region 
of the fundamental plane and $\kappa-$space.
All of these massive clusters have comparable masses, 
mass-to-light ratios, and central surface brightnesses. 

Support of the interpretation of these clusters as tidally-stripped 
galaxy nuclei is provided by the tentative detection of extratidal light 
by \citet{harris02} for six of the globular clusters (marked with an $x$ in 
Table~\ref{tbl:params} and \ref{tbl:mass}, {\it partly filled circles} in the 
figures). 
In the context of the stripped-dwarf model, it is tempting to view this 
extratidal light as the last vestiges of the extended dwarf envelope around 
these nuclei, although extratidal light may also be present due to the 
evaporation of cluster stars by two-body relaxation. 
Deeper observations of these and other massive clusters 
would be extremely valuable to confirm and quantify this extratidal light. 
Such observations would also better quantify the ellipticities of these 
globular clusters, which are comparable to those of $\omega$Cen and G1 but 
greater than those of dwarf nuclei. 

These globular clusters may also simply represent the upper end of the 
globular cluster mass function in \cena. The similarities between these 
globular clusters and the most massive globular clusters in the Local 
Group are more established than the interpretation of the better-studied 
Local Group clusters as tidally-stripped dwarf nuclei. 
In addition, there is evidence 
for yet more massive young star clusters that demonstrate that there is 
overlap between the masses of the most massive star clusters and the 
least massive dwarf galaxies. 
The most massive star cluster known is W3, a cluster with $\sigma = 45$ \kms\  
in the merger remnant NGC~7252 \citep{maraston04}. The inferred mass of this 
cluster is nearly $10^8$ \msun, even more massive than the 
nuclei of the dwarf galaxies studied by \citet{geha02}. 
Even if we were to make the extreme suggestion that all of the \cena\ globular 
clusters are in fact relic dwarf nuclei, the existence of W3 demonstrates 
there is overlap between the masses of galaxies and star clusters. 

On the other side of the mass gap, the ultra-compact dwarf galaxies in the 
Fornax cluster \citep{drinkwater03} are actually {\it less} massive than W3 
and overlap with the most massive \cena\ clusters. These dwarfs have velocity 
dispersions of $\sigma = 24 - 37$ \kms, effective radii of 10 -- 30 pc, 
masses in the range $10^{7-8}$ \msun, and mass-to-light ratios of 2 -- 4 in 
solar units \citep{drinkwater03}. They are thus very comparable to the larger 
and more massive of the \cena\ globular clusters. Numerical models for these 
ultra-compact dwarfs have shown that they can form from nucleated dwarfs that 
have been stripped of their envelopes by tidal forces in a cluster 
\citep{bekki03a}. This stripping process can explain the ultra-compact 
dwarfs in Fornax, although it will operate over approximately a factor of 
2 smaller radius in a smaller group like \cena\ due to decreased tidal shear. 
The radius for the tidal stripping of \cena\ globular cluster progenitors is 
likely to be on order 10~kpc \citep[][see their Figure~7]{bekki03a}, comparable 
to the projected distances of 5 -- 23~kpc for some of the \cena\ globular 
clusters in this study. It is therefore plausible that tidal stripping 
could have transformed at least some nucleated dwarf galaxies into 
these \cena\ globular clusters. 

Another aspect of the tidally-stripped dwarf model is that their dark matter
halos cannot be too cuspy as otherwise the relatively concentrated 
dark matter core will be too effective at retaining the stellar envelope.  
In their study, \citet{bekki03a} used the dark matter profile of 
\citet{salucci00} (originally proposed by \citet{burkert95} for 
dwarf galaxies), rather than the more commonly adopted 
profile of \citet*{navarro96}, because it has a flatter core. 
However, this requirement for a relatively flat dark matter core may remove 
one possible explanation for the higher mass-to-light ratio of these massive 
globular clusters. While a residual dark matter halo from their dwarf galaxy 
past is a possible explanation for the larger mass-to-light ratios of these 
\cena\ clusters, a dark matter profile with a flat, low-density core will 
also be more efficiently stripped away \citep{bekki03a}. 
An alternative explanation for the high mass-to-light ratios of these
massive globular clusters is if they formed in a starburst with a 
truncated stellar initial mass function \citep{charlot93}. A mass function with 
relatively more low-mass stars would produce a higher mass-to-light 
ratio. 

The detailed shapes and kinematics of the nuclei of nucleated dwarfs and 
the most massive globular clusters may provide one way to further investigate 
the potential connection between these two populations. The most massive 
globular clusters have significant ellipticities \citep{harris02}, while
this does not appear to be the case for the nuclei of nucleated dwarfs 
\citep{geha02}. 
However, the same tidal forces that strip a dwarf envelope may also induce 
significant ellipticities. 
The importance of rotational flattening and anisotropies 
in the velocity distribution may also serve to distinguish between 
dwarf nuclei and globular clusters. 

The hypothesis that some of the most massive globular clusters are the nuclei 
of galaxies offers an appealing explanation for recent evidence of an 
intermediate-mass black hole in G1 \citep*{gebhardt02}. The mass of this 
black hole falls on the same $M_{BH} - \sigma$ relationship for galaxies 
and suggests that the formation mechanisms for black holes in star clusters 
and galaxy spheroids are similar. If G1 is instead simply a tidally-stripped, 
nucleated dwarf 
galaxy, the problem is reduced in complexity and only one physical mechanism 
for black hole growth is required to explain the $M_{BH} - \sigma$ relation. 

An alternate interpretation of the similarity between the most massive 
globular clusters and the nuclei of nucleated dwarfs is that the later are 
simply star clusters that have migrated to or formed at the centers of these 
dwarfs. 
The properties of these objects would then be probes of massive star clusters 
in different environments, rather than of actual overlap between the properties 
of the most massive star clusters and the least massive galaxies. 
This interpretation also explains their similar location in $\kappa-$space, 
although stands in contrast to the simple explanation for the 
intermediate-mass black hole in G1. 

\section{Summary} \label{sec:sum}

We have measured velocity dispersions for a sample of 14 
globular clusters in the nearby, luminous elliptical galaxy \cena, 
the first such study of the globular cluster system of a luminous 
elliptical galaxy and the first such study outside the Local Group. 
These clusters have velocity dispersions in the range 10 -- 30 \kms, 
comparable to the largest previously measured values for globular clusters. 
We have used measured King model structural parameters for these clusters 
from the literature to derive masses for all 14 
clusters. These clusters are comparable in mass to the most massive 
Galactic globular cluster $\omega$Cen and M31's G1. 
From these data we find the following: 

\noindent
1. The globular clusters in \cena\ approximately follow the same 
fundamental plane relationships as Local Group globular clusters and extend 
them to approximately an order of magnitude higher mass and luminosity. 

\noindent
2. The mean mass-to-light ratio of these clusters is larger than for 
typical Local Group globular clusters, although comparable
to the more massive Local Group clusters. 

\noindent
3. These clusters begin to bridge the mass gap between the most massive 
globular clusters and the least massive dwarf galaxies. In particular, there 
is very good overlap in the photometric, structural, and kinematic 
properties of these clusters and the properties of both nucleated dwarf 
elliptical nuclei and ultra-compact dwarf galaxies. 

\noindent 
4. The large masses of these clusters, combined with the possible detection 
of extratidal light for some objects by \citet{harris02}, suggest that 
some of these globular clusters are in fact the nuclei of tidally stripped 
dwarf galaxies.

The common properties of the most massive star clusters and the nuclei of 
the least massive dwarfs suggest that both the formation mechanisms for 
star cluster and galaxies can produce objects in the same region of the 
fundamental plane or $\kappa-$space. 
Alternately, the nuclei of nucleated dwarf galaxies may simply be star clusters 
that happen to lie in the centers of galaxies, rather than true galaxy nuclei. 
Future spectroscopic observations of additional massive 
globular clusters could quantify the relative contribution of relic dwarf 
nuclei to this population through kinematics and with stellar population 
models, while deep, high-resolution images could provide better measurements 
of structural parameters, particularly in the core, and search for and study 
diffuse, low surface-brightness envelopes. 

\acknowledgements 

We would like to thank the staff of Las Campanas Observatory for their 
excellent support, in particular for making the MIKE spectrograph 
available after domestic security concerns delayed PANIC. 
We acknowledge useful discussions with Dan Kelson and thank Francois 
Schweizer, John Huchra, and the referee for helpful comments on the manuscript. 
PM received support from a Carnegie Starr Fellowship and a Clay Fellowship. 
The research of LCH is supported by the Carnegie Institution of Washington and
by NASA grants from the Space Telescope Science Institute (operated by AURA,
Inc., under NASA contract NAS5-26555).

\bibliographystyle{apj}
%\bibliography{$HOME/tex/references/references}

\end{document}